\pdfoutput=1
\documentclass[sigconf]{acmart}

\usepackage{enumitem}
\usepackage{xspace}
\usepackage{arydshln}
\usepackage{multirow}

\newcommand{\ignore}[1]{}

\newcommand{\daat}{\textsc{DaaT}\xspace}
\newcommand{\saat}{\textsc{SaaT}\xspace}
\newcommand{\bmw}{\textsf{BMW}\xspace}
\newcommand{\vbmw}{\textsf{VBMW}\xspace}
\newcommand{\wand}{\textsf{WAND}\xspace}
\newcommand{\maxscore}{\textsf{MaxScore}\xspace}

\newcommand{\an}[1]{$^{_{_{^{^{#1}}}}}$}

\setcopyright{none}
\renewcommand\footnotetextcopyrightpermission[1]{}

\begin{document}

\title{Wacky Weights in Learned Sparse Representations and the Revenge of Score-at-a-Time Query Evaluation}

\author{Joel Mackenzie,\an{1} Andrew Trotman,\an{2} Jimmy Lin\an{3}}

\affiliation{\vspace{0.1cm}
$^1$ School of Computing and Information Systems, The University of Melbourne \country{Australia} \\
$^2$ Department of Computer Science, University of Otago, Dunedin, New Zealand \\
$^3$ David R. Cheriton School of Computer Science, University of Waterloo, Ontario \country{Canada} \\
}

\begin{abstract}
Recent advances in retrieval models based on learned sparse representations generated by transformers have led us to, once again, consider score-at-a-time query evaluation techniques for the top-$k$ retrieval problem.
Previous studies comparing document-at-a-time and score-at-a-time approaches have consistently found that the former approach yields lower mean query latency, although the latter approach has more predictable query latency.
In our experiments with four different retrieval models that exploit representational learning with bags of words, we find that transformers generate ``wacky weights'' that appear to greatly reduce the opportunities for skipping and early exiting optimizations that lie at the core of standard document-at-a-time techniques.
As a result, score-at-a-time approaches appear to be more competitive in terms of query evaluation latency than in previous studies.
We find that, if an effectiveness loss of up to three percent can be tolerated, a score-at-a-time approach can yield substantial gains in mean query latency while at the same time dramatically reducing tail latency.
\end{abstract}

\renewcommand{\shortauthors}{}
\pagestyle{empty}

\maketitle

\section{Introduction}

Despite various investigations of alternatives over the past few decades, document-at-a-time (\daat) query evaluation algorithms remain the dominant solution for the top-$k$ retrieval problem that lies at the core of systems-focused information retrieval research as well as deployed production systems.
The most recent systematic exploration of different query evaluation strategies that we are aware of is the work of Crane et al.~\cite{Crane_etal_WSDM2017}, who compared the then latest score-at-a-time (\saat) query evaluation algorithm against the {\wand}-family of \daat index-traversal techniques~\cite{bc+03-cikm, Ding_Suel_SIGIR2011}.
They concluded that despite advances in score-at-a-time query evaluation, the best approach at the time, known as JASS~\cite{Lin_Trotman_ICTIR2015,Trotman_Crane_2019}, was still slower than Block-Max {\wand} (\bmw) in terms of mean query latency, although the query latency of JASS was more consistent, with much lower tail latency~\cite{Dean_Barroso_CACM2013}.

In this work, we once again compare \daat and \saat query evaluation techniques, but in the context of a new class of retrieval models that rely on learned sparse representations based on transformers.
A natural question to ask:\ Why are we relitigating these comparisons?
Recent empirical studies have shown that this new class of retrieval models yields effectiveness that is superior to unsupervised sparse retrieval models such as BM25, and at least on par with popular dense retrieval techniques~\cite{Lin_arXiv2021_repir}.
Thus, it is important to study the behavior of query evaluation techniques in the context of these models.

Interestingly, we observe that the neural models (typically, transformers) that underlie these learned sparse retrieval models often assign ``wacky'' weights to terms in bag-of-words representations.
The adjective ``wacky'' is used in at least two ways:\
First, the distribution of term weights does not appear to allow standard \daat optimizations to effectively perform skipping and early exiting, which represent the source of accelerated query evaluation performance with traditional bag-of-words scoring models like BM25.
Second, manual examination of the assigned term weights reveals that high scores are frequently assigned to terms in counter-intuitive ways, e.g., large weights placed on stopwords and subwords that lack any meaningful semantic content.
These weight assignments are particularly puzzling since the models have been demonstrated to be more effective than existing bag-of-words models such as BM25 on various benchmark datasets.

The contributions of this work are twofold:\ First, we systematically characterize the wackiness of these learned sparse representations and their effects on query evaluation performance.  
Although these issues have been observed before~\cite{Mallia_etal_SIGIR2021,Lin_etal_DESIRES2021}, our work confirms that these wacky weights are indeed quite pervasive.
Second, we are the first to compare \daat vs.\ \saat approaches in this context and show that this wackiness differentially affects \daat more than \saat.
More precisely, the weight distributions in learned sparse representations appear to greatly reduce opportunities for skipping and early exiting, thus reducing the efficiency of \daat approaches.
In contrast, since \saat approaches rely less on advantageous weight distributions, slowdown is less pronounced.
On the whole, if approximate query evaluation with an effectiveness loss of up to three percent can be tolerated, a score-at-a-time approach can yield substantial gains in mean query latency while at the same time dramatically reducing tail latency.

\section{Background and Related Work}

The standard formulation of document retrieval using bag-of-words representations can be distilled into the following scoring function:
\begin{equation}
S_{d,q} = \sum_{t \in d \cap q} w_{d,t} \cdot w_{q,t}
\label{scoring}
\end{equation}
\noindent where $w_{d,t}$ is the weight of term $t$ in document $d$ and $w_{q,t}$ represents the weight of term $t$ in the query $q$.
Document weights are typically computed via a function of term statistics such as tf, idf, doclength, etc.
Query weights are often set to one, which simplifies query--document scores to the sum of document weights of terms that are found in the query.
This formulation covers nearly all major families of retrieval models (probabilistic, vector space, language modeling, divergence from randomness, etc.) and is equivalent to the inner product of two weighted vectors of dimension $|V|$, where $V$ is the vocabulary of the collection.

How to efficiently generate a top-$k$ ranking of documents from an arbitrarily large collection $\mathcal{C} = \{ d_i \}_{i=0}^{N}$ in terms of $S_{d,q}$ has been, quite literally, the subject of many decades of intense study; see Tonellotto et al.~\cite{Tonellotto_etal_FnTIR2018} for a survey.
In the IR context, this is often referred to as the query evaluation problem, and nearly all modern algorithms exploit an inverted index where postings lists are systematically visited (i.e., traversed) in order to generate the top-$k$ documents, stored in data structures such as min-heaps, accumulator tables, or a combination thereof.

Document-at-a-time (\daat) and score-at-a-time (\saat) are apt descriptions of the different approaches that query evaluation algorithms can take.
At a high level, \daat approaches work on postings lists that are monotonically sorted by document identifier, and achieve low latency query evaluation by ``avoiding work''.
Through the use of auxiliary data structures and per-term score upper-bounds, algorithms can cleverly work out when one or more documents cannot possibly be in the top-$k$ and thereby ``skip'' them (often en masse).
Presently, Variable Block-Max {\wand} (\vbmw)~\cite{Mallia_etal_SIGIR2017} is generally acknowledged to represent the state of the art, although the best choice of \daat traversal algorithm can depend on properties of the collection, query stream, ranker, and specific optimizations enabled~\cite{khe20-sigir, cikm20mm, mss19-ecir, pcm13-adcs}.

In contrast, \saat approaches depend on term weights that have first been quantized into integers (called impact scores) and organized in postings lists grouped by descending impact scores~\cite{akm01-sigir}.
Query evaluation proceeds by considering the score contributions of terms, from highest to lowest.
Instead of ``avoiding work'', the intuition behind \saat is to organize computations to maximally take advantage of modern processors architectures (e.g., avoiding pipeline stalls, cache misses, etc.).
The most recent take on \saat query evaluation is the technique known as JASS~\cite{Lin_Trotman_ICTIR2015,Trotman_Crane_2019}, which has the additional advantage of being an ``anytime algorithm'':\ since term contributions are considered in decreasing order of importance (i.e., contributions to the final score), query evaluation can terminate at any time to yield an approximate ranking.

The experiments of Crane et al.~\cite{Crane_etal_WSDM2017} showed, on a number of standard IR test collections, that \bmw was faster than JASS (i.e., lower mean query latency), but is susceptible to higher tail latency.
That is, for some relatively small fraction of queries, ill-behaved term weights rendered \bmw much slower than JASS.
This behavior had been previously noted by Petri et al.~\cite{pcm13-adcs}, who examined the efficiency of both {\wand} and {\bmw} under different bag-of-words ranking models.
Subsequent work demonstrated that both {\daat} and {\saat} techniques can benefit from additional optimizations such as document identifier reassignment {\cite{tois22mpm, mss19-ecir}}.

The performance benefits of the various optimization techniques described above depend naturally on the scoring function.
For example, the ability to skip blocks of postings depends on the relationship between the various term weights, and how they are distributed across the index (for block-based pruning approaches).
Previous evaluations have been based on ``traditional'' scoring function such as BM25 or language models~\cite{pcm13-adcs}, but recently the field has seen the emergence of models where term weights are {\it learned} (so-called learned sparse representations~\cite{Lin_Ma_arXiv2021,Lin_arXiv2021_repir}).
These models still rely on bag-of-words representations, and thus top-$k$ retrieval can still be captured by Eq.~(\ref{scoring}), and all the foregoing discussions about \daat vs.~\saat approaches still apply.
However, term weights are now supplied by neural networks (today, transformer models), and learned in a supervised manner from large amounts of training data.

The first example of this class of models using transformers is DeepCT~\cite{Dai:1910.10687:2019}, which used a regression model to learn term weights.
Mackenize et al.~\cite{Mackenize_etal_SIGIR2020} showed that these weights affect the behavior of \daat query evaluation approaches, although, with appropriate mitigation, it is possible to accelerate query evaluation without harming effectiveness.
In other words, the distribution of term weights assigned by the neural model is qualitatively different from BM25.
Additional evidence for this finding comes from the work of Mallia et al.~\cite{Mallia_etal_SIGIR2021} in the context of their proposed model called DeepImpact:\ query latency is noticeably longer with term weights assigned by a transformer model, compared to the BM25 score over the same sets of terms in each document.
To further compound this issue, some learned models also assign weights to {\emph{queries}}, thereby changing the relative importance between the terms and potentially hindering efficiency further.
These interesting observations suggest that we should take another detailed look at query evaluation algorithms in the context learned sparse retrieval models, and thus the ``previously settled'' \daat vs.\ \saat debate should be reopened.

Note that our work only examines learned {\it sparse} representations for retrieval, where documents are represented by bags of words, i.e., the basis of the vector representation is the vocabulary space.
There is, of course, another large class of learned {\it dense} representations, exemplified by models such as DPR~\cite{karpukhin-etal-2020-dense} and ANCE~\cite{Xiong_etal_ICLR2021}; see Lin et al.~\cite{Lin_etal_2021_ptr4tr} for a survey.
Although transformer-based models are involved in both classes of retrieval models, dense retrieval techniques require a completely different ``software stack''.
For example, top-$k$ retrieval is formulated as nearest neighbor search and implemented with approximate techniques such as HNSW~\cite{Malkov_Yashunin_2020}.
These techniques do not use inverted indexes and thus our discussions of \daat and \saat query evaluation algorithms are not applicable.
However, for the interested reader, a number of papers have attempted to draw conceptual connections between dense and sparse learned representations~\cite{Lin_Ma_arXiv2021,Lin_etal_DESIRES2021,Lin_arXiv2021_repir}.

\section{Experimental Setup}

Our experiments used the popular MS MARCO passage corpus, comprising 8.8M passages, and all models were evaluated on the 6980 queries in the development set~\cite{MS_MARCO_v3}.
While it would have been desirable to explore multiple test collections, we are limited by the trained models that are publicly available for download, since training models from scratch is beyond the scope of this work.
We hope to examine more test collections in future work.

\subsection{Retrieval Models}

As points of comparison, we adopted the following baselines:

\smallskip
\noindent {\bf BM25} simply performs retrieval using the ubiquitous BM25 scoring function~\cite{Robertson_Zaragoza_FnTIR2009} over bag-of-words representations of the passages in the corpus.
We set $k_1 = 0.82$ and $b = 0.68$, based on tuning on a selection of training instances on the MS MARCO passage ranking test collection~\cite{Lin_etal_SIGIR2021_Pyserini}.

\smallskip
\noindent {\bf BM25 w/ doc2query--T5}~\cite{Nogueira_etal_arXiv2019,Nogueira_Lin_docTTTTTquery} (BM25-T5 for short) augments passages in the corpus with query predictions generated by the T5~\cite{Raffel_etal_JMLR2020} neural sequence-to-sequence model.
The expanded passages are scored using BM25 at retrieval time, with the same BM25 formulation and parameters above. 
Thus, while neural models are involved in corpus preparation, the assignment of term weights does not involve any neural networks.

\smallskip
\noindent
We examined the following retrieval models that leverage sparse learned representations using transformers:

\smallskip
\noindent {\bf DeepImpact}~\cite{Mallia_etal_SIGIR2021} uses doc2query--T5 to identify dimensions in the passage's bag-of-words representation that should have non-zero weights (i.e., expansion terms) and learns a term weighting model based on a pairwise loss between relevant and non-relevant passages with respect to a query.

\smallskip
\noindent {\bf uniCOIL + doc2query--T5}~\cite{Lin_Ma_arXiv2021} (uniCOIL-T5 for short) is a simplified variant of COIL~\cite{Gao_etal_ECIR2021} that assigns scalar weights to terms (as opposed to vector weights in the original COIL formulation).
This model additionally benefits from doc2query--T5 expansions.

\smallskip
\noindent {\bf uniCOIL + TILDE}~\cite{Zhuang:2108.08513:2021} (uniCOIL-TILDE for short) can be best characterized as replacing the doc2query--T5 expansion component with an alternative model based on TILDE~\citep{Zhuang_Zuccon_SIGIR2021} that has lower inference costs but appears to be just as effective.

\smallskip
\noindent {\bf SPLADEv2}~\cite{Formal:2109.10086:2021} represents an improvement over SPLADEv1~\citep{Formal_etal_SIGIR2021}, which itself builds on SparTerm~\citep{Bai:2010.00768:2020}.
For this family of sparse retrieval models, the expansion component can be best characterized as being based on masked language modeling.
SPLADEv2 further improves effectiveness via distillation techniques.

\medskip
\noindent
Note that for all the models, our experiments are based on our own implementation of code and generation of data, with Anserini as a starting point (see details below).
In all applicable cases, we started with checkpoints of the neural models provided by the authors.
Thus, our results are quite close, but not exactly the same, as figures reported in the respective authors' original papers.
For the sparse learned retrieval models, the corpus and queries are both pre-tokenized.
The corpus already includes term weights for each term, and the same for the queries.
Thus, none of these experiments involved neural inference, which eliminates a source of non-determinism in neural models.

\subsection{Systems}
\label{systems}

We conducted experiments with three different retrieval systems:

\smallskip
\noindent {\bf Anserini}~\cite{Yang_etal_JDIQ2018} is an IR toolkit built on the open-source Lucene search library written in Java.
The version of Anserini used in our experiments is based on Lucene 8.3.0, which uses \bmw for query evaluation~\cite{Grand_etal_ECIR2020}.
We made no special modifications to the default query evaluation settings.

\smallskip
\noindent {\bf PISA}~\cite{Mallia_etal_2019} is an efficiency-focused IR system written in C++.
In our experiments, PISA executes {\sf{MaxScore}} {\cite{tf95-ipm}} processing over {\sf{SIMD-BP128}} encoded postings lists~\cite{lb15-spe}. 
While {\vbmw} is currently the state-of-the-art approach for \daat traversal, recent work has demonstrated that {\maxscore} is a better choice for larger values of $k$ (such as $k = 1000$) and for long queries.

\smallskip
\noindent {\bf JASS}~\cite{Lin_Trotman_ICTIR2015,Trotman_Crane_2019} is, as far as we are aware, the only actively maintained open-source \textsc{SaaT} search engine available today.
It, too, was implemented in C++ for efficiency and uses a SIMD-enhanced Elias gamma encoding scheme for compressing the postings~\cite{Trotman_Lilly_2018}.
Query evaluation proceeds by processing impact-ordered postings using simple integer arithmetic and storing the results in an accumulator array managed with a unique page-table like algorithm.
JASS controls the tradeoff between query latency and effectiveness through a parameter $\rho$, which limits the total number of postings processed per query.
Although JASS usually employs 16-bit accumulators (allowing maximum per document scores of $2^{16}$), 32-bit accumulators were necessary in these experiments in order to avoid overflows, as the learned sparse impacts and weights often result in scores exceeding $2^{16} = 65{,}536$.
A preliminary benchmark suggests that moving from 16 to 32-bit accumulators adds an almost $50$\% overhead to the exact (rank-safe) BoW BM25 run in Table~\ref{result:msmarco-passage}, row (3a).

\medskip
\noindent
Our implementations used the ``pseudo-document'' trick to assign custom impact scores for each document term.
That is, if term {\it X} was assigned a (quantized) integer weight of ten by the transformer, we simply repeat the term ten times in a dynamically created ``fake'' document.
This allows us to use all the existing systems without any modifications, by simply swapping in ``sum of term frequency'' as the scoring function. 
Note, however, that both \daat systems {\emph{do not}} pre-quantize the BM25 or BM25-T5 indexes, computing the scores on the fly instead.

Finally, all experiments were conducted in memory using a single thread on a Linux machine with two 3.50 GHz Intel Xeon Gold 6144 CPUs and 512 GiB of RAM.
Indexes were built with Anserini and exported to the Common Index File Format~\cite{lm+20-sigir} before being imported into PISA and JASS.
In addition, both PISA and JASS made use of document reordering~\cite{Mackenzie_etal_SIGIR2021, dk+16-kdd} which has been shown to improve compression and accelerate query processing in both \daat and \saat ~\cite{tois22mpm}. 
In our experiments, all systems retrieved the top $k = 1000$ documents. 

\begin{table}[t]
\centering\scalebox{0.87}{
\begin{tabular}{llrrr}
\toprule
 & & \multicolumn{1}{c}{\textbf{Quality}} & \multicolumn{1}{c}{\textbf{Time}} & \multicolumn{1}{c}{\textbf{Space}} \\
 \cmidrule(lr){3-5}
 \multicolumn{2}{l}{\bf Method} & RR@10 & Latency & Index Size \\
   &  &  & (ms) & (MB) \\
\toprule
\multicolumn{2}{l}{\bf Anserini (Lucene): \daat} \\
(1a) & BM25 & 0.187 & 40.1 & 661\\
(1b) & BM25-T5 & 0.277 & 62.8 & 1036\\
(1c) & DeepImpact & 0.325 & 244.1 & 1417\\
(1d) & uniCOIL-T5 & 0.352 &  222.3 & 1313 \\
(1e) & uniCOIL-TILDE & 0.350 & 194.6 & 2067\\
(1f) & SPLADEv2 & 0.369 & 2140.0 & 4987 \\
\midrule
\multicolumn{2}{l}{\bf PISA: \daat} \\
(2a) & BM25 & 0.187 & 8.3 & 739\\
(2b) & BM25-T5 & 0.276 & 11.9 & 1150\\
(2c) & DeepImpact & 0.326 & 19.4 & 1564\\
(2d) & uniCOIL-T5 & 0.352 & 36.9 & 1358 \\
(2e) & uniCOIL-TILDE & 0.350 & 28.4 & 2108 \\
(2f) & SPLADEv2 & 0.369 & 220.3 & 4326\\
\midrule
\multicolumn{2}{l}{\bf JASS: \saat} \\
\multicolumn{2}{l}{\it Exact} \\
(3a) & BM25 &  0.187 & 15.8 & 1156\\
(3b) & BM25-T5 & 0.277 & 50.2 & 1452\\
(3c) & DeepImpact & 0.326 & 39.3 & 2039 \\
(3d) & uniCOIL-T5 & 0.352 & 147.2 & 1310 \\
(3e) & uniCOIL-TILDE &  0.350 & 83.5 & 1976 \\
(3f) & SPLADEv2 &  0.369 & 314.0 & 3813\\
[0.3ex]
\cdashline{1-5}\noalign{\vskip 0.5ex}
\multicolumn{2}{l}{\it Approximate} \\
(4a) & BM25 & 0.186 & 12.4 & 1156\\
(4b) & BM25-T5 & 0.275 & 10.1 & 1452\\
(4c) & DeepImpact & 0.319 & 12.6 & 2039 \\
(4d) & uniCOIL-T5 & 0.338 & 14.9 & 1310 \\
(4e) & uniCOIL-TILDE & 0.338 & 15.4 & 1976 \\
(4f) & SPLADEv2 & 0.324 & 15.3 & 3813\\
\bottomrule
\end{tabular}
}
\vspace{0.25cm}
\caption{Experimental results on the development queries of the MS MARCO passage ranking test collection. 
\label{result:msmarco-passage}
}
\vspace{-0.4cm}
\end{table}

\section{Results}

\subsection{Main Findings}

Our main results are presented in Table~\ref{result:msmarco-passage}, where each combination of retrieval model and system is characterized in terms of output quality, time, and space.
Output quality is measured in terms of mean reciprocal rank at cutoff 10 (RR@10), the official metric of the test collection; time is measured in terms of query latency in milliseconds, and space in terms of index size measured in megabytes.
Note that latency figures do not include the time taken to encode or expand the queries for the learned sparse retrieval models since we used pre-tokenized queries with pre-computed weights.
While PISA and JASS were designed specifically as platforms for research in query evaluation algorithms and thus engineered for speed, none of the systems had small index sizes explicitly as a design goal (beyond incidental effects on query evaluation performance that result from index compression techniques).

Each block of the table is devoted to a system, and rows represent each retrieval model.
Note that JASS results occupy two blocks; the top one represents {\emph{exact}} (i.e., rank-safe) processing (where all postings are processed) and the bottom one represents {\emph{approximate}} processing.
For the latter, we used a maximum of $\rho = 1$ million postings visited, based on heuristics provided by the JASS authors~\cite{Lin_Trotman_ICTIR2015,Trotman_Crane_2019}.

There are a few takeaways from these results:
First, our experiments confirm what previous researchers have already discovered~\cite{Mackenize_etal_SIGIR2020,Mallia_etal_SIGIR2021}---that retrieval models based on learned sparse representations substantially alter the behavior of query evaluation algorithms.
However, our results more thoroughly and comprehensively capture the effects, illustrating that these issues pervade sparse learned models more broadly.
For \daat approaches, both Anserini (Lucene) and PISA, we see that, in general, models with higher effectiveness are slower.
In Lucene, the difference between BoW BM25 in row (1a) and the most effective model, SPLADEv2 in row (1f) is a whopping $\sim$50$\times$ slowdown.
With PISA, the slowdown is ``only'' $\sim$25$\times$, but nevertheless still quite dramatic.
Even for models that are slightly less effective than SPLADEv2, for example, the uniCOIL models in rows (d) and (e), we observe substantially worse query evaluation performance, in both Lucene and PISA.

Second, it is clear that the query evaluation performance of Anserini (Lucene) is far behind that of PISA, both in absolute terms as well as relative degradation with the models that we have explored.
Of course, this is not a fair comparison because Lucene is production-ready search library that is already widely deployed, whereas PISA is a research system.
The performance deficiencies of Lucene have been discussed elsewhere~\cite{Grand_etal_ECIR2020} so we don't beleaguer the point further, so for the remainder of this paper we focus our comparisons on PISA.

To further examine different query evaluation algorithms, we applied {\wand} and {\bmw} to SPLADEv2 in PISA in a side experiment and found that these algorithms resulted in {\emph{slower}} processing than an {\emph{exhaustive ranked disjunction}}, with mean latency values of 619ms, 681ms, and 553ms, respectively. In our experiments, {\maxscore} vastly outperforms the {\wand}-based approaches; this result is due to {\maxscore} avoiding expensive sorting operations during query processing {\cite{mss19-ecir}}.
Essentially, {\wand} and {\bmw} are ``working hard'' to compute which documents can be skipped, but the work is essentially wasted because in reality, few documents can actually be skipped if exact query evaluation is desired.
This, once again, shows that procrastination pays.

Third, the JASS results from Table~\ref{result:msmarco-passage} are largely consistent with what we already know about \saat approaches from the literature.
Exact query evaluation (i.e., exhaustively traversing all postings) with JASS is slower than PISA, but achieves comparable effectiveness.
For the learned sparse retrieval models, although we do observe performance degradation with JASS as well, the slowdowns are slightly less than what we observed with PISA; ``only'' around $20\times$ when comparing BoW BM25 (3a) with SPLADEv2 (3f).

The approximate query evaluation results with JASS are also consistent with previous findings.
For ``traditional'' BM25 weighting, in rows (4a) and (4b), JASS is able to speed up query evaluation at the cost of small decreases in effectiveness.
However, as the ``model sophistication'' increases, the heuristics provided by the JASS authors appear to result in an increased loss in effectiveness.
Looking at rows (4d) to (4f), while JASS is much faster than PISA, it comes at a drop in effectiveness. 
Although the loss is less than 1\% for BM25, and 4\% on DeepImpact and UniCOIL, on SPLADEv2, the loss is over 12\%, which can be attributed to the smaller fraction of total postings being processed~\cite{msc17-adcs}.
Table~\ref{result:msmarco-passage} only captures two operating points for JASS, but we explore additional configurations in Section~\ref{sec-to} that yield different effectiveness/efficiency tradeoffs.

Finally, in terms of index sizes, there's a general trend of larger index sizes as effectiveness increases; the underlying reasons will become obvious in the next section.
However, for all models and systems, the space requirements are quite modest on the whole.
In the context of modern servers, where terabytes of disk storage are common, these differences are negligible.

\subsection{Wacky Weights}
\label{section:wacky}

\begin{table}
\centering
\scalebox{0.87}{
\begin{tabular}{ l r r r r r}
\toprule
\multirow{2}{*}{Method} & \multicolumn{1}{c}{\multirow{2}{*}{$|V|$}} & \multicolumn{2}{c}{Terms in Documents} & \multicolumn{2}{c}{Terms in Queries}\\ 
\cmidrule(lr){3-4}\cmidrule(lr){5-6}
&& Total & Unique & Total & Unique\\
\midrule
BM25          & 2660824 & 39.8    & 30.1  & 5.9    & 5.8 \\
BM25-T5       & 3929111 & 224.7   & 51.1  & 5.9    & 5.8 \\
DeepImpact    & 3514102 & 4010.0  & 71.1  & 4.2    & 4.2 \\
uniCOIL-T5    & 27678   & 5032.3  & 66.4  & 686.3  & 6.6 \\
uniCOIL-TILDE & 27646   & 8260.8  & 107.6 & 661.1  & 6.5 \\
SPLADEv2      & 28131   & 10794.8 & 229.4 & 2037.8 & 25.0 \\
\bottomrule
\end{tabular}
}
\vspace{0.25cm}
\caption{Term statistics of documents and queries for the different treatments of the MS MARCO passage corpus.
\label{term-stats}
}
\vspace{-0.4cm}
\end{table}


What might be the cause of the behavior described in the previous section?
Table~\ref{term-stats} begins to answer some of these questions.
The two main sections in the table show descriptive statistics of the documents and queries.
An explanation about the total vs.\ unique terms:\ the total number of terms includes duplicated terms in our pseudo-documents, as described in Section~\ref{systems}.
Thus, total terms is best understood as the sum of all the weights assigned to all unique terms, either in the document or the query.
The second column, $|V|$, denotes the vocabulary size, or the total number of unique terms in the collection; this is also the number of dimensions that are in the representation vectors of the documents and queries.

A few additional caveats are necessary for properly understanding these figures.
All statistics are computed based on our replication of the retrieval models, and thus may differ from figures reported in the authors' original papers due to differences in corpus preparation.
In all cases, counts are computed by simply splitting text on whitespace.
For BM25 and BM25-T5, these are {\it prior} to tokenization, stopword removal, etc.
For uniCOIL and SPLADEv2, both document and queries are pre-tokenized.
Thus, there are qualitative differences, especially since uniCOIL and SPLADEv2 tokens are subwords derived from BERT, e.g., ``androgen receptor'' is broken into ``and \#\#rogen receptor''.
Thus, for this reason, uniCOIL and SPLADEv2 have much smaller vocabulary sizes.

Nevertheless, these statistics go a long way to explaining many of the observations from Table~\ref{result:msmarco-passage}.
It is clear that learned sparse retrieval models achieve higher effectiveness based on document expansion, and in some models, query expansion.
Document expansion obviously yields longer documents (and bigger indexes), and together with query expansion, it is no mystery why query evaluation efficiency degrades.
These models, however, begin to give us a more nuanced look.
For example, comparing DeepImpact and uniCOIL-T5, we see that the latter performs query expansion, whereas the former doesn't.
This is likely the biggest source of query latency differences, row (2c) vs.\ row (2d) in Table~\ref{result:msmarco-passage}.
Comparing the T5 and TILDE variants of uniCOIL, we observe that they achieve comparable effectiveness, but T5 is more ``compact'' in performing less document expansion.
However, counter-intuitively, uniCOIL-TILDE, row (2e), is actually faster than uniCOIL-T5, row (2d).
We don't presently have a good explanation for this.
Finally, we note that SPLADEv2 likely derives its superior effectiveness from performing more expansion.
Compared to the uniCOIL models, the SPLADEv2 queries contain around $4\times$ more unique terms and the documents contain $2-4\times$ more unique terms.

\begin{figure}
\centering
\includegraphics[width=.95\columnwidth]{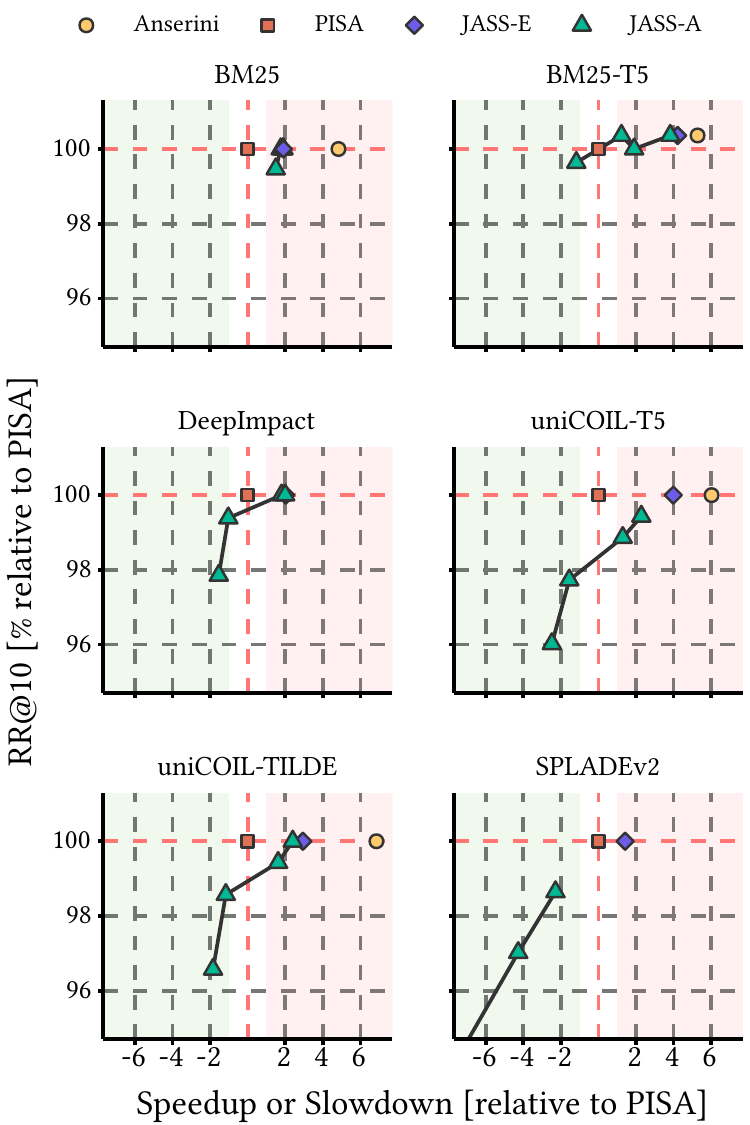}
\caption{Relative comparisons between each system configuration for each retrieval model. The $x$-axis reports the relative speedup (negative values) or slowdown (positive values) with respect to PISA; the $y$-axis reports the relative effectiveness as a percentage of mean RR@10 achieved by PISA.
\label{fig-tradeoff-delta}
}
\end{figure}

\begin{figure*}[t]
\centering
\includegraphics[width=.9\textwidth]{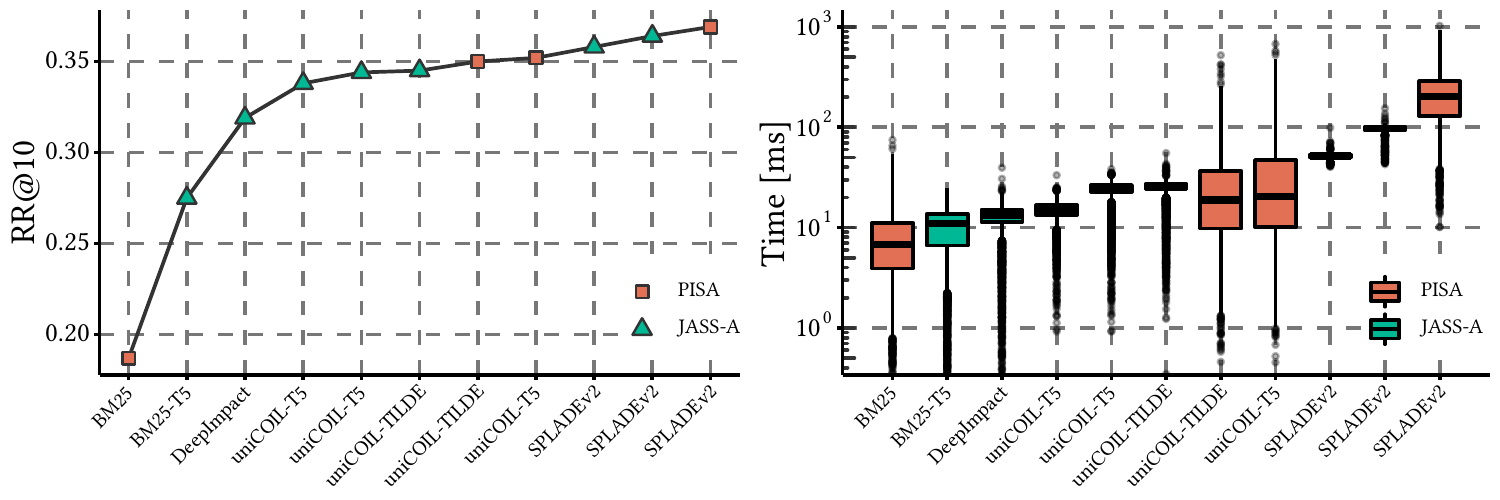}
\caption{Effectiveness (left) and distribution of query latency (right) for all configurations along the Pareto-optimal frontier from Figure~\ref{fig-tradeoff}. \daat \maxscore in PISA exhibits much greater variability compared to the \saat approach in JASS.
\label{fig-pareto-compare}
}
\end{figure*}

\begin{figure}
\centering
\includegraphics[width=.95\columnwidth]{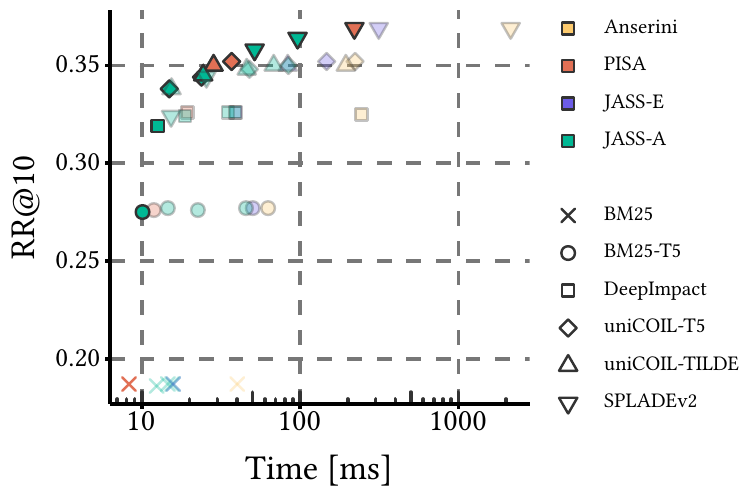}
\caption{Efficiency (mean query latency) vs.\ effectiveness (mean RR@10) for all configurations; those along the Pareto-optimal frontier are highlighted. Note that {\emph{every}} retrieval model is Pareto-optimal under at least one configuration for either PISA or JASS. However, Lucene does not appear on the frontier.
\label{fig-tradeoff}
}
\end{figure}




The wackiness of term weights assigned by learned sparse retrieval models is evident based on manual examination of model output.
Building on the tokenization example above, for the query ``androgen receptor define'', the full SPLADEv2 query includes 25 unique tokens, representing an increase of 21 tokens beyond those in the original query, ``and \#\#rogen receptor define''.
For these tokens, the model assigns the following weights: (``and'', 225), (``\#\#rogen'', 251), (``receptor'', 242), and (``define'': 59).
These weights seem reasonable, although the large weight on ``and'' is a bit surprising, given that the subword is conflated with a stopword.
Many expansion terms that SPLADEv2 adds to the query do make sense, for example, (``hormone'', 179), (``definition'', 162), and (``meaning'', 99).
However, puzzling is the fact that the model also adds many stopwords to the query, including (``is'', 70), (``the'', 56), (``for'', 46), and (``are'', 32).
Most non-sensical is the fact that ``,'' (yes, the comma) is added as an expansion term, with a relatively large weight of 68 (srsly, wtf?).
Note that as far as we are aware, SPLADEv2 was trained without any special corpus processing, and punctuation are indeed part of BERT's vocabulary.
In some cases, punctuation are semantically meaningful (e.g., apostrophes in names like O'Toole), but it is difficult to see the role that the comma might play in the context of this query.
While it is difficult to argue with the overall effectiveness results (the high mean RR@10), we nevertheless find these weights ``wacky''.

\subsection{Effectiveness/Efficiency Tradeoffs}
\label{sec-to}

One main advantage of \saat approaches, and JASS specifically, is the ability to trade effectiveness for efficiency by processing varying amounts of postings.
Since the query evaluation algorithm considers segments of postings in decreasing order of importance, it is possible to quit at any time, yielding approximate results.
These tradeoffs are shown in Figure~\ref{fig-tradeoff-delta}, with a plot for each model.

In each plot, we have normalized both the effectiveness and efficiency of the PISA configuration to one, represented as a red square at (1, 100), and thus both effectiveness (in percent) and latency speedup/slowdown are in relative terms (negative $x$ represents speedup, positive $x$ represents slowdown).
Lucene is represented as a yellow circle; it is just as effective as PISA but substantially slower.
JASS with exhaustive query evaluation (JASS-E) is shown as a blue diamond; in general, its effectiveness is comparable to PISA and Lucene, while being slower than PISA but faster than Lucene.
The green triangles capture the tradeoff curve of JASS with approximate query evaluation using $\rho \in \{1, 2, 5, 10\}$ million postings; in the case of SPLADEv2, some points are outside the plot area.

By altering $\rho$ we can trade effectiveness for efficiency to differing degrees, beyond the two operating points reported in Table \ref{scoring}.
For example, considering the most effective model, SPLADEv2, the loss in mean RR@10 is less than $2$\% with a mean query latency of $96$ms (half the time of PISA) with $\rho = 10$m.
If even more effectiveness loss is tolerable, using $\rho = 5$m can yield a mean query latency of $52$ms (quarter of the time of PISA) with a $3$\% loss in mean RR@10. 
On the other hand, PISA is a better choice if no effectiveness loss is acceptable, regardless of the retrieval model. 

In Figure~\ref{fig-tradeoff} we have combined all six subplots in Figure~\ref{fig-tradeoff-delta} into a single plot.
Here, we report absolute measures of effectiveness (mean RR@10) as well as efficiency (mean query latency).
This plot captures all combinations of retrieval models (as shapes) and systems (as colors).
Combinations that are on the Pareto-optimal frontier are highlighted.

In this context, the Pareto-optimal frontier represents the best tradeoff that can be obtained for all explored combinations of retrieval models $\times$ systems.
If a point lies on the frontier, it means that no other setting is able to achieve {\it both} higher effectiveness and lower mean query latency.
There is no principled way to identify ``better'' or ``worse'' configurations along the frontier, as the selection of the operating point depends on the application scenario.
Points along the frontier represent that best of ``what's possible'' for the system designer to choose from.

Quite amazingly, we note that {\it all} models lie on some part of the frontier.
This means that there is no single model that dominates all others.
Furthermore, PISA and JASS-A (approximate query evaluation) share points along the frontier.
This means that depending on the desired tradeoffs between effectiveness and efficiency, the optimal choice varies, both in terms of the retrieval model as well as the query evaluation approach!

Finally, Figure~\ref{fig-pareto-compare} provides more details on the configurations along the Pareto-optimal frontier.
In the left plot, we have arranged the configurations along the $x$ axis to clearly show each model/system combination, with the corresponding mean RR@10 shown along the $y$-axis.
Here, we can clearly see that each retrieval model represents an optimal choice given some desired level of effectiveness.
In the right plot, we show the distribution of query latency as standard Tukey box-and-whiskers plots; it is important to note that the selection of the Pareto-optimal configurations was performed using {\it mean} query latency, and means might differ from medians due to outliers.

Confirming prior work~\cite{Crane_etal_WSDM2017}, we observe that while the \daat  query evaluation can outperform \saat query evaluation in terms of mean latency, better performance comes at the expense of less predictable latency, i.e., tail latency.
On the other hand, \saat query evaluation, by design, yields much more predictable latency, as it enforces a strict limit on the total amount of allocated computation on a per query basis. 
It is also worth noting the lack of variance in the latency of JASS with the SPLADEv2 model.
Since SPLADEv2 aggressively expands queries (see Table~\ref{term-stats}), JASS almost always processes $\rho$ postings, resulting in highly predictable performance.
In contrast, for some of the other models, queries may have fewer than $\rho$ postings to process, resulting in the ``lower tails'' (i.e., queries that are much faster), for example, in JASS with BM25-T5.

\section{Conclusions}

Retrieval models based on sparse learned representations are relatively new innovations.
Most work has focused on evaluating model effectiveness, but here we build on previous studies to examine effectiveness/efficiency tradeoffs.
We find that, indeed, term weights generated by such models have substantive effects on the behavior of query processing algorithms, and that they differentially affect \daat vs.\ \saat.
The net effect is to make \saat algorithms more attractive in general.

In future work, it would be interesting to include the recently proposed {\emph{anytime}} \daat algorithms to the comparison~\cite{tois22mpm}, as they provide similar work-limiting (and, in turn, tail-latency minimizing) guarantees as \saat retrieval.
Furthermore, expanding our experimental comparisons to different collections would help us understand corpus effects.
Finally, an obvious improvement for retrieval models based on learned sparse representations would be to incorporate efficiency considerations into the model training objective, much like the ``learning to efficiently rank'' thread of work from a decade ago~\cite{Wang_etal_SIGIR2010,Wang_etal_SIGIR2011}.
How to formulate an appropriate loss, however, is not obvious.
Nevertheless, there are many exciting directions we are interested in pursuing in the near future.

\section*{Acknowledgements}

This research was partially supported by the Australian Research Council Discovery Project DP200103136 and the Natural Sciences and Engineering Research Council (NSERC) of Canada.
We thank Antonio Mallia, Alistair Moffat, and Matthias Petri for their helpful discussions related to this work.


\bibliographystyle{abbrv}
\bibliography{sample-base}

\end{document}